\documentclass[%
 aip,
 amsmath,amssymb,
 reprint,%
]{revtex4-1}

\usepackage{graphicx}
\usepackage{dcolumn}
\usepackage{bm}

\usepackage{amsmath}
\usepackage{amssymb}
\usepackage{graphicx}
\usepackage{indentfirst}
\usepackage{booktabs}
\usepackage{multirow}
\usepackage{colortbl}

\usepackage[utf8]{inputenc}
\usepackage[T1]{fontenc}
\usepackage{mathptmx}
\usepackage{etoolbox}

\makeatletter
\def\@email#1#2{%
 \endgroup
 \patchcmd{\titleblock@produce}
  {\frontmatter@RRAPformat}
  {\frontmatter@RRAPformat{\produce@RRAP{*#1\href{mailto:#2}{#2}}}\frontmatter@RRAPformat}
  {}{}
}%
\makeatother
\begin{document}

\preprint{AIP/123-QED}

\title{Sliding ferroelectricity tunable conventional and anomalous spin Hall effects in bilayer 1T'-WTe$_2$}

\author{Chao Wu}
\affiliation{State Key Laboratory for Mechanical Behavior of Materials, School of Materials Science and Engineering, Xi'an Jiaotong University, Xi'an, Shaanxi, 710049, People's Republic of China}
\affiliation{Shaanxi Yanyi Titanium Industry Company Limited, Baoji, Shaanxi, 721001, People's Republic of China}
\author{Pengqiang Dong}
\affiliation{State Key Laboratory for Mechanical Behavior of Materials, School of Materials Science and Engineering, Xi'an Jiaotong University, Xi'an, Shaanxi, 710049, People's Republic of China}
\affiliation{Shaanxi Yanyi Titanium Industry Company Limited, Baoji, Shaanxi, 721001, People's Republic of China}
\author{Kai Wei}
\affiliation{State Key Laboratory for Mechanical Behavior of Materials, School of Materials Science and Engineering, Xi'an Jiaotong University, Xi'an, Shaanxi, 710049, People's Republic of China}
\affiliation{Shaanxi Yanyi Titanium Industry Company Limited, Baoji, Shaanxi, 721001, People's Republic of China}
\author{Hanbo Sun}
\affiliation{State Key Laboratory for Mechanical Behavior of Materials, School of Materials Science and Engineering, Xi'an Jiaotong University, Xi'an, Shaanxi, 710049, People's Republic of China}
\affiliation{Shaanxi Yanyi Titanium Industry Company Limited, Baoji, Shaanxi, 721001, People's Republic of China}
\author{Ping Li}
\email{pli@xjtu.edu.cn}
\altaffiliation{Email: pli@xjtu.edu.cn}
\affiliation{State Key Laboratory for Mechanical Behavior of Materials, School of Materials Science and Engineering, Xi'an Jiaotong University, Xi'an, Shaanxi, 710049, People's Republic of China}
\affiliation{Shaanxi Yanyi Titanium Industry Company Limited, Baoji, Shaanxi, 721001, People's Republic of China}
\affiliation{National Laboratory of Solid State Microstructures, Nanjing University, Nanjing, Jiangsu, 210093, People's Republic of China}
\affiliation{Hefei National Research Center for Physical Sciences at the Microscale, University of Science and Technology of China, Hefei, Anhui, 230026, People's Republic of China}

\date{\today}

\begin{abstract}
The spin Hall effect, recognized for its high-speed, low-power, and highly controllable characteristics, is a key enabler for next-generation memory and logic devices. However, a primary challenge lies in achieving 180$^{\circ}$ magnetization switching without an external magnetic field in spin-orbit torque devices. Here, we propose a method to tune the conventional and anomalous spin Hall effects by the intrinsic sliding ferroelectricity. Importantly, the anomalous spin Hall effect can enable the field-free switching of perpendicular magnetization. We find a substantial anomalous spin Hall conductivity of $\sigma_{xy}^{y}$ = 45.62 ($\hbar$/e)S/cm and $\sigma_{yx}^{y}$ = 56.84 ($\hbar$/e)S/cm in monolayer 1T'-WTe$_2$. These values are significantly enhanced to $\sigma_{xy}^{y}$ = -96.77 ($\hbar$/e)S/cm and $\sigma_{yx}^{y}$ = 104.03 ($\hbar$/e)S/cm in the bilayer 1T'-WTe$_2$. More interestingly, the sliding ferroelectricity enables reversible switching of the signs and magnitudes for both the conventional and anomalous spin Hall conductivities. This originates from the fact that the sliding ferroelectric markedly shifts the relative spin Berry curvature contributions from the valence and conduction bands around the $\Gamma$-X path. Our findings not only reveal a strong coupling between sliding ferroelectricity and spin transport, but also propose a strategy for the nonvolatile electrical control of spintronic devices.
\end{abstract}

\maketitle

The spin Hall effect (SHE) is defined as the generation of a transverse spin-polarized current by an applied longitudinal electric field, resulting in opposite spin accumulations at the edges of the material \cite{1,2,3,4}. Spin current has attracted widespread attention due to its ability to carry and process information, making it an integral component of spintronics \cite{5,6,7,8,9}. When a strong spin-orbit interaction (SOI) exists within a material, the carrier current $J$ flowing along the $\hat{j}$ direction generates a spin current $J_i^k$, where i and k denote the spin current direction $\hat{i}$ and spin polarization direction $\hat{k}$, respectively. Therefore, within the Cartesian coordinate, the spin current is expressed as $J_i^k=\theta_0\hbar/\left(2e\right)\epsilon_{ijk}J_k$, where $\epsilon_{ijk}$ denotes the Levi-Civita symbol and the Einstein summation convention is implied. $\theta_0$ is a material parameter known as the spin Hall angle, whose exact value is currently debated. In conventional spin Hall effects (CSHE), the direction vectors $\hat{i}$, $\hat{k}$, and the carrier current direction $\hat{j}$ is orthogonal to each other. Recently, Wang \cite{10} suggested that an anomalous spin Hall effect (ASHE) can arise from the interaction of conducting electrons with order parameters, such as the magnetization $\textbf{M}$ in ferromagnets and the $\rm N\acute{e}el$ order in antiferromagnets. Namely, the $\hat{i}$, $\hat{j}$, $\hat{k}$ is not orthogonal to each other. So, one of the great advantages of ASHE is that one can control the generated spin current by controlling the magnetization or the $N\acute{e}el$ order in the magnetic materials. The ASHE is not limited to magnetic materials. In fact, it suggests that any order parameter breaking inversion symmetry can generate an anomalous spin Hall conductivity (ASHC) through the SOI \cite{11}. The generation of pure spin currents via the CSHE and ASHE in the absence of an external magnetic field offers a highly efficient route for electrically controlling electron spins \cite{3,12}. This spin current can produce an out-of-plane anti-damping spin-orbit torque (SOT), which is essential for achieving field-free switching of perpendicular magnetization \cite{13,14,15,16}. Currently, many experiments have already shown that this effect can realize field-free switching of perpendicular magnetization \cite{13,16,17,18,19,20,21}. Researchers achieved field-free switching of perpendicular magnetization in systems TaIrTe$_4$, IrMn, and WTe$_2$ by reducing the lattice symmetry \cite{13,16,17,18,19,21}. Moreover, the non-collinear antiferromagnetic order in the Mn$_3$Ir system can also achieve field-free switching of perpendicular magnetization \cite{20}. Although these findings underscore the application potential of both the CSHE and ASHE, efficiently and controllably tuning these effects remains a significant challenge.

\begin{figure*}[htb]
\begin{center}
\includegraphics[angle=0,width=0.8\linewidth]{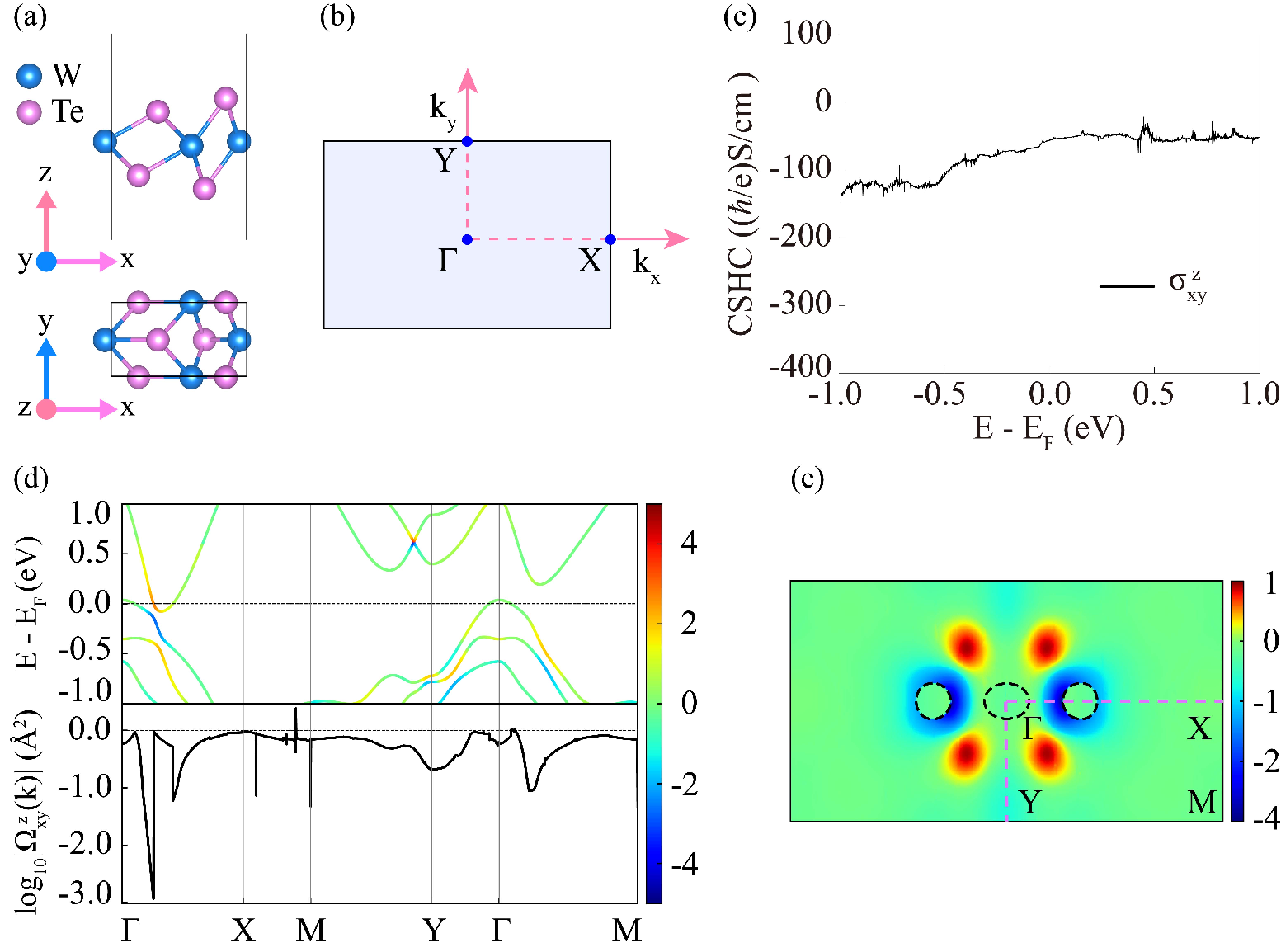}
\caption{(a) The atomic structure of monolayer 1T'-WTe$_2$. (b) Schematic diagram of the 2D BZ of 1T'-WTe$_2$, the high-symmetry points are indicated by blue dots. (c) The CSHC $\sigma_{xy}^z$ of monolayer 1T'-WTe$_2$ as a function of energy. (d) The upper panel shows the band structure projected by the SBC on a log scale, and the lower panel shows the k-resolved SBC at E-E$_F$ = 0 eV. (e) The k-resolved SBC on a log scale in 2D BZ slice of k$_z$ = 0 for monolayer 1T'-WTe$_2$ at E-E$_F$ = 0 eV. The high-symmetry points are $\Gamma$(0, 0, 0), X(0.5, 0, 0), Y(0, 0.5, 0), and M(0.5, 0.5, 0). The black dashed lines are the intersections of the Fermi surface with a slice of the BZ.}
\end{center}
\end{figure*}

Sliding ferroelectricity emerges in bilayer or multilayer systems from symmetry breaking in specific stacking configurations, generating a reversible out-of-plane polarization via interlayer sliding \cite{22,23,24,25}. Sliding ferroelectric possess ultralow switching barriers and relatively high Curie temperatures, which underpin their capabilities for fast, energy-efficient, and fatigue-resistant information writing. These advantageous properties have been firmly established by experimental studies \cite{26,27,28,29}. Furthermore, the integration of sliding ferroelectric with the diverse properties of two-dimensional (2D) materials provides a powerful means to control material functionality via sliding. Recent theoretical studies have highlighted the versatile coupling between sliding ferroelectricity and diverse electronic phenomena. For instance, Huang $\emph{et al.}$ predicted a coupling with nonlinear Hall effects in ZrTe$_5$ thin layers \cite{30}. Yang $\emph{et al.}$ demonstrated that sliding ferroelectricity can drive a phase transition from antiferromagnetism to altermagnetism in SnS$_2$/MnPSe$_3$/SnS$_2$ heterostructure \cite{31}. Expanding the scope to multiferroic functionality, Li $\emph{et al.}$ proposed that sliding ferroelectricity can be coupled with magnetic phase transitions, valley polarization reversal, and topological phase transitions \cite{32,33,34,35}. Given its proven ability to control fundamental magnetic and valley properties, a key question is whether sliding ferroelectricity can also effectively tune the CSHE and ASHE.

In this article, we propose that sliding ferroelectricity can tune not only the magnitude of the ASHC and the conventional SHC (CSHC), but also their orientation. Firstly, we find that monolayer 1T'-WTe$_2$ has a larger CSHC ($\sigma_{xy}^{z}$ = -55.98 ($\hbar$/e)S/cm, $\sigma_{yx}^{z}$ = 48.46 ($\hbar$/e)S/cm) and ASHC ($\sigma_{xy}^{y}$ = 45.62 ($\hbar$/e)S/cm, $\sigma_{yx}^{y}$ = 56.84 ($\hbar$/e)S/cm) at the Fermi level. Then, all bilayer stackings exhibit a nearly twice increase in the CSHC. However, the ASHC of the AA stacking also increased by nearly twice, but the ASHC of the AB stackings decreased significantly. Moreover, switchable ferroelectricity is observed only for sliding along the x-axis, with sliding along the y-axis resulting merely in a net polarization. The sliding ferroelectricity is $6.98\times 10^{\text{-}13}$ \text{C/m}. When the ferroelectric polarization is tuned by interlayer sliding, the CSHC and ASHC evolve correspondingly. The interlayer siliding transition from the paraelectric phase AB-0 stacking, through the ferroelectric phase AB-1 stacking, to the AB-2 stacking, the CSHCs values remained relatively stable, while the ASHCs values initially increased and then decreased. We reveal the microscopic mechanism underlying the variation of the CSHCs and ASHCs, which stem from the change in the spin Berry curvature (SBC) of the valence and conduction bands along the $\Gamma$-X path. Our work not only establishes a strong coupling between sliding ferroelectricity and the CSHE and ASHE, but also proposes a strategy for the non-volatile electrical control of spin currents. This approach provides a novel pathway for developing next-generation, low-power spintronic devices.

\begin{figure*}[htb]
\begin{center}
\includegraphics[angle=0,width=0.8\linewidth]{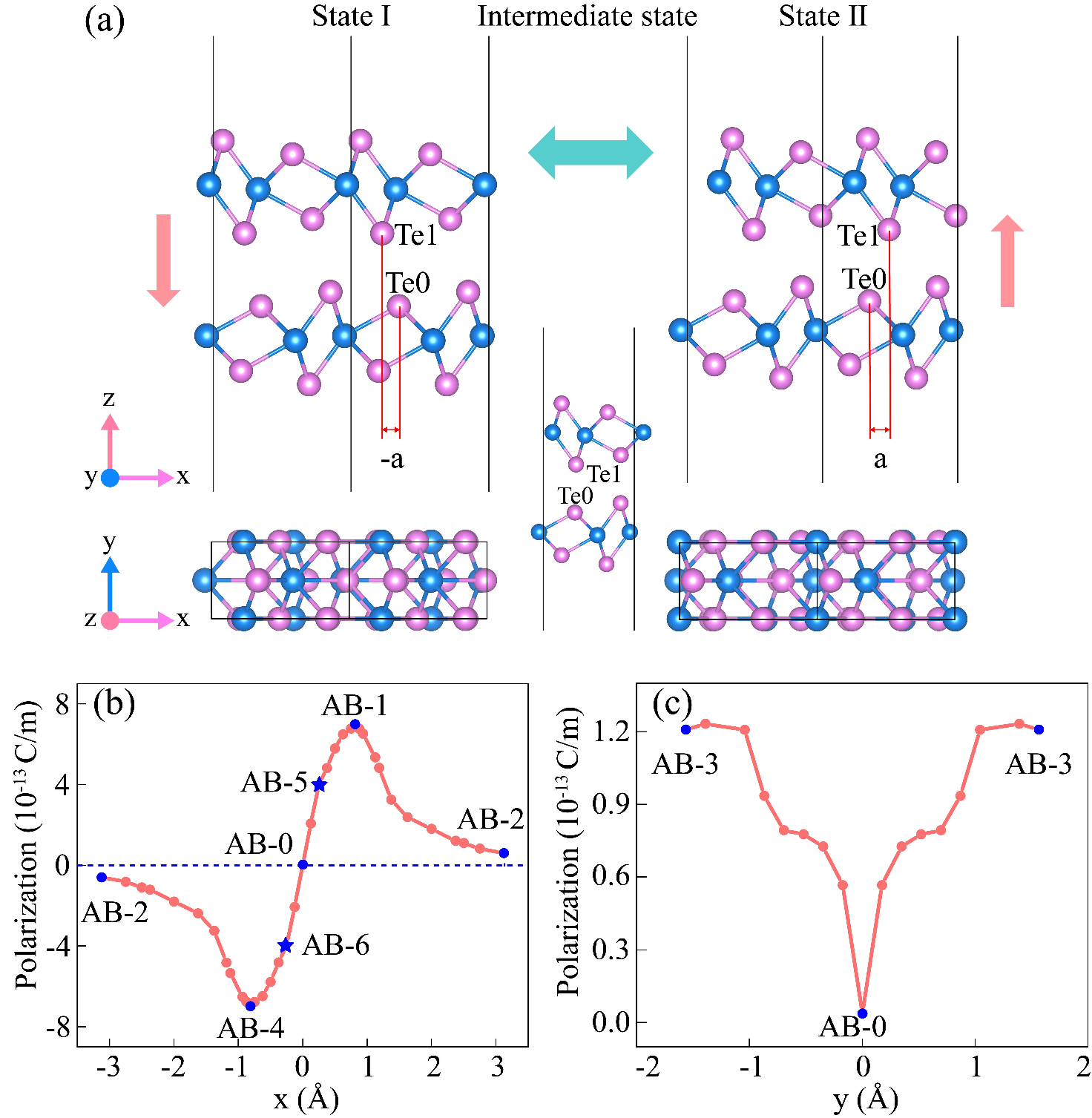}
\caption{(a) The top and side views of the crystal structure for state I bilayer 1T'-WTe$_2$. The state II is the reflecting state I by the central horizontal plane. They are the AB-5 and AB-6 stackings. (b) Ferroelectric switching pathway along the x-axis with respect to the slip x distance at y = 0 {\AA}. (c) Ferroelectric switching pathway along the y-axis with respect to the slip y distance at x = 0 {\AA}. The orange colour arrows represent the direction of ferroelectric polarization.}
\end{center}
\end{figure*}

Our density functional theory (DFT) calculations are carried out using the Vienna ab initio simulation package (VASP). The projector augmented wave (PAW) potentials and the generalized gradient approximation (GGA) in the Perdew-Burke-Ernzerhof (PBE) \cite{36,37,38,39} form are applied. The plane-wave cutoff energy is set to 500 eV. The zero damping DFT-D3 method is used to account for van der Waals (vdW) interaction \cite{40}. The $\Gamma$-centered k-mesh of $11 \times 7 \times 1$ and $24 \times 12 \times 1$ are used for structural optimization and self-consistent calculations and a vacuum spacing of $\sim$19 {\AA} is adopted. The calculations employed force criteria of 0.001 eV/{\AA}. The vertical polarization is assessed with dipole moment corrections. The switching pathway for sliding ferroelectricity is obtained via the nudged elastic band (NEB) method \cite{41}. The DFT wave functions are converted into a maximally localized Wannier functions (MLWFs) using WANNIER90 software \cite{42,43}, and the CSHCs and ASHCs are calculated employing the Kubo formula. We performed K-mesh tests on the CSHC of monolayer and bilayer 1T'-WTe$_2$. The $450\times450\times1$ and $150\times150\times1$ K-mesh have converged for monolayer and bilayer 1T'-WTe$_2$ (see Fig. S1 and Fig. S2). A dense k-mesh of $450\times450\times1$ (monolayer) and $150\times150\times1$ (bilayer) is used for integrate over the BZ to calculate the intrinsic CSHCs and ASHCs, and the rapid variation of SBC is addressed through adaptive k-mesh refinement. The Kubo formula of the intrinsic CSHCs and ASHCs are expressed as \cite{44,45,46,47}:
\begin{equation}
\sigma_{\alpha \beta}^{\gamma}=-\frac{e^{2}}{\hbar}\frac{1}{VN_{ k}^{3}}\sum_{ \scriptsize\bm{k}}\Omega_{\alpha \beta}^{\gamma}( \bm{k})
\end{equation}
where the k resolved term is:
\begin{equation}
\Omega^{\gamma}_{\alpha \beta}( \bm{k})=\sum_{n}f_{n \scriptsize\bm{k }}\Omega^{\gamma}_{n,\alpha \beta}( \bm{k})
\end{equation} The $\alpha$, $\beta$, and $\gamma$ represent the directions of the spin current, the applied electrical field, and the spin polarization of the spin current, respectively. And the SBC term can be written is:
\begin{equation}
\Omega^{\gamma}_{n,\alpha \beta}( \bm{k}) = \hbar^{2}\sum_{m\neq n}\frac{-2 {\rm Im}\left[\langle n \bm{k}|\frac{1}{2}\{\hat{\sigma}_{\gamma},\hat{\bm{v}}_{\alpha}\}|m \bm{k} \rangle\langle m \bm{k}|\hat{\bm{v}}_{\beta}|n \bm{k} \rangle\right]}{(\epsilon_{n \scriptsize\bm{k}}-\epsilon_{m \scriptsize\bm{k}})^{2}}
\end{equation}
where V, $N_k^3$, $f_{n\bm{k}}$, ${\hat{v}}_x$ and ${\hat{v}}_y$ are the primitive cell volume, the number of k points in BZ, the Fermi distribution function, and the velocity operator, respectively.

\begin{table*}[htb]
\caption{
The calculated CSHC and ASHC for monolayer 1T'-WTe$_2$, four typical AB stackings and AA stacking bilayer 1T'-WTe$_2$. The CSHC and ASHC are represented by $\sigma^{\gamma}_{
\alpha\beta}$, where the indices $\alpha$, $\beta$, and $\gamma$ correspond to the directions of the spin current, the applied electrical field, and the spin polarization of the spin current, respectively (unit: ($\hbar$/e)S/cm). }
	\resizebox{1.5\columnwidth}{!}{	
	\begin{tabular}{ccccccccccccc}
		\hline
		System          & SHC                   & Monolayer      & AB-0       & AB-1      & AB-2     & AB-3     & AB-4     & AB-5     & AB-6     & AA      \\
		\hline
		               &$\sigma_{xy}^{z}$      & -56.13         & -121.10    & -128.09   & -116.98  & -108.13  & -128.19  & -125.46  & -125.46  & -118.21  \\
		\raisebox{1ex} {Conventional}
		                &$\sigma_{yx}^{z}$      & 48.62          & 64.65      & 75.79     & 85.23    & 61.15    & 75.91    & 65.22    & 65.62    & 74.03   \\
		\hline
	                	&$\sigma_{xy}^{y}$      & -45.71         & -0.12      & -10.91    & 2.68     & -9.34    & 10.93    & -2.48    & 2.42     & -96.77  \\
		Anomalous       &$\sigma_{yx}^{y}$      & 56.98          & 5.19       & 14.75     & 4.17     & 5.95     & -14.83   &  6.29    & -6.02    & 104.03  \\
                 		&$\sigma_{yy}^{x}$      & 2.76           & 0.50       & 8.12      & -1.92    & 2.66     & -8.08    & -0.23    & 0.26     & 0.09    \\
		\hline
	\end{tabular}}
\end{table*}	

\begin{figure*}[htb]
\begin{center}
\includegraphics[angle=0,width=0.8\linewidth]{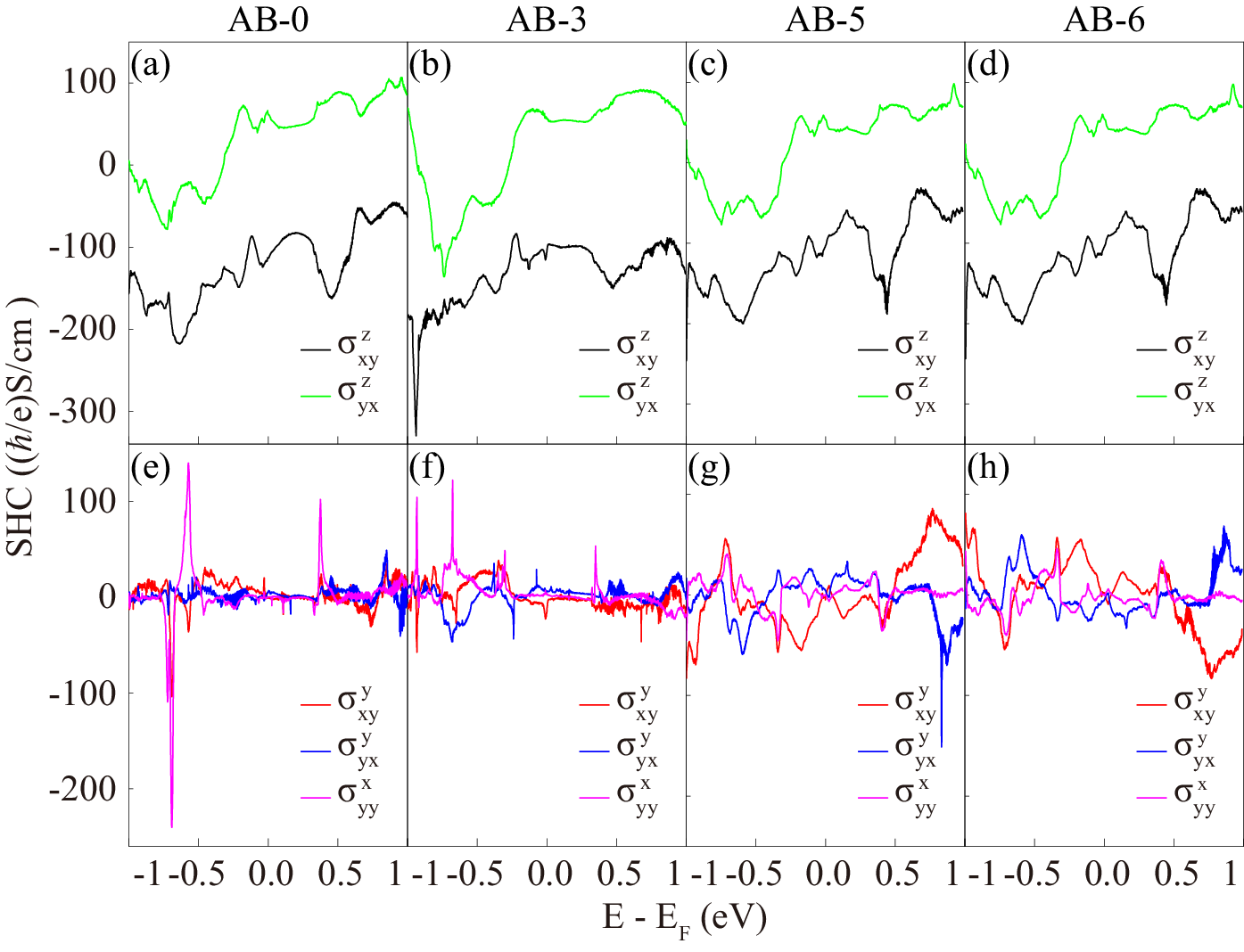}
\caption{The (a-d) CSHC and (e-h) ASHC tensor components as a function of energy for the four AB stacking configurations.}
\end{center}
\end{figure*}

\begin{figure*}[htb]
\begin{center}
\includegraphics[angle=0,width=1.0\linewidth]{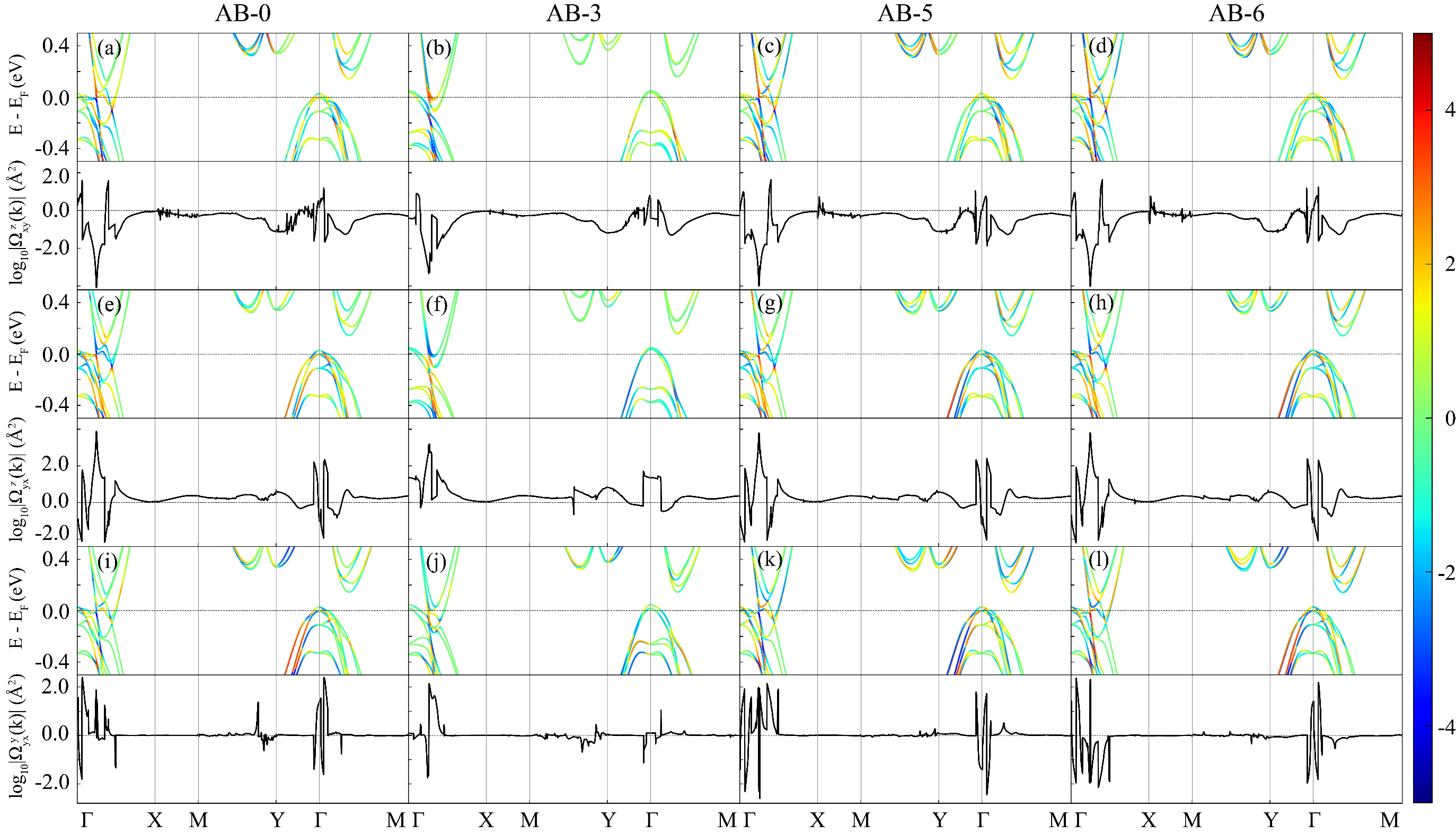}
\caption{The band structures and SBCs for different AB stacking structures. The AB-0, AB-3, AB-5, and AB-6 stacking configurations correspond to (a, e, i), (b, f, j), (c, g, k), and (d, h, l), respectively. The upper panel shows the band structure projected by the SBC on a log scale, and the lower panel shows the k-resolved SBC at E-E$_F$ = 0 eV for (a-d)$\sigma_{xy}^{z}$, (e-h)$\sigma_{yx}^{z}$, (i-l)$\sigma_{yx}^{y}$, respectively.
}
\end{center}
\end{figure*}

As shown in Fig. 1(a), monolayer 1T'-WTe$_2$ is an orthogonal transition metal dichalcogenide (TMD) with strong spin-orbit coupling (SOC) \cite{48}. The monolayer structure belong to the polar orthorhombic space group Pmn2$_1$ (No. 31). The monolayer 1T'-WTe$_2$ crystal structure has a mirror symmetry M$_x$. The optimized lattice constant is a = 3.68 {\AA} and b = 6.25 {\AA}. Fig. 1(b) exhibits the Brillouin zone (BZ) and high-symmetry point. Since the monolayer 1T'-WTe$_2$ is a 2D system, the concepts of spin current and charge current along the z-axis are not meaningful. Therefore, only the $\sigma_{xy}^y$, $\sigma_{yx}^y$, and $\sigma_{yy}^x$ components of the ASHC remain \cite{11}. For 2D materials, the calculate CSHC and ASHC using \cite{49,50}
\begin{equation}
\sigma_{\alpha \beta}^{\gamma '} = \sigma_{\alpha \beta}^{\gamma} \times \frac{l_z}{l^0_z}
\end{equation}
where $\sigma_{\alpha \beta}^{\gamma}$ is the original CSHC or ASHC value, l$_z$ and l$^0_z$ are the height along z-axis including the vacuum layer and only 1T'-WTe$_2$, respectively. Table I exhibit the calculated CSHCs and ASHCs at Fermi level. The metal Pt exhibits $\sigma_{xy}^z$ = -$\sigma_{yx}^z$ due to the C$_{4v}$ symmetry \cite{51}. The lowered symmetry of monolayer 1T'-WTe$_2$ structure results in pronounced anisotropy in the CSHCs. The $\sigma_{xy}^z$ is -55.98 ($\hbar$/e)S/cm , while the $\sigma_{yx}^z$ is 48.46 ($\hbar$/e)S/cm. As shown in Fig. 1(c), we calculate the CSHC tensor element $\sigma_{xy}^z$ for monolayer 1T'-WTe$_2$. To clarify the origin of CSHCs and ASHCs, we calculate the band structure and k-resolved SBC. Here, we take $\sigma_{xy}^z$ at Fermi level as an example. The upper panel of Fig. 1(d) exhibits the band structure of the SBC $\Omega_{n,xy}^z(\bm{k})$ projected on the log scale \cite{52,53}, where the blue(red) colour represents the negative(positive) contribution to the SBC. The lower panel of Fig. 1(d) shows the k-resolved SBC on the log scale of monolayer 1T'-WTe$_2$. The SBC exhibits strong variations along the $\Gamma$-X and Y-M paths, with particularly pronounced peaks along $\Gamma$-X. These peaks arise from band crossings near the Fermi level. Compared to the band structure without spin-orbit coupling (SOC) (Fig. S3), the SOC-induced band splitting along the $\Gamma$-X path generates a negative SBC near the Fermi level. To comprehensively understand the distribution of the SBC throughout the BZ, we calculate the k-resolved SBC distribution at k$_z$ = 0 in the 2D BZ. As shown in Fig. 1(e), it can be clearly observed that there are two negative contributions on the $\Gamma$-X path, while four positive contributions are present near the $\Gamma$-M path. Consequently, the integrated negative contribution surpasses the positive one, leading to a net negative SBC that accounts for the overall negative $\sigma_{xy}^z$. The analysis above confirms that the SOC-induced band gap near the Fermi level will significantly affect the SHC.

To understand the effect of sliding ferroelectricity on the CSHC and ASHC of bilayer 1T'-WTe$_2$, we investigates the evolution of the sliding path and the electronic structure. We investigate two sliding pathways by exploring displacements along the x- and y-axes. It should be noted that sliding along the y-axis cannot realize the transition from state I to state II, since the AB stacking bilayer 1T'-WTe$_2$ also possesses M$_x$ mirror symmetry. Therefore, the focus is primarily on sliding along the x-axis. Starting from the non-polar intermediate state (see Fig. 2(a)), the state I (state II) can be obtained by translating the upper layer along the -x (+x) axis. Namely, when the upper Te1 atoms shifted -a (+a) relative to the lower Te0 atoms, the system transforms into the ferroelectric polarization downwards (ferroelectric polarization upwards) phase from the paraelectric phase.

In order to understand the stable configuration of interlayer stacking, as show in Fig. S4, we investigated the sliding energy barriers along x-axis and y-axis. When the interlayer of bilayer 1T'-WTe$_2$ slides along the x-axis, we observe an interesting phenomenon. The AB-1 and AB-4 stackings with the largest ferroelectric polarization values do not correspond to the minimum stacking energy (see Fig. 2(b)). The most stable stacking configurations are AB-5 and AB-6, which are 25.98 meV lower in energy than the AB-1 and AB-4 stackings with the highest polarization values. The AB-5 and AB-6 stackings are 0.64 meV lower in energy than the paraelectric AB-0 stacking. When sliding along the y-axis, as shown in Fig. S4 (c), the paraelectric phase has the lowest energy. Fig. 2(b) displays the evolution of the polarization with the slip distance x along the x-axis. When the slip distance varies, switchable electrostatic polarization emerges, demonstrating sliding ferroelectricity. The AB-0 stacking is non-polar intermediate state. When the slip of 0.25 {\AA} becomes an AB-5 stacking configuration, it exhibits a ferroelectric polarization of $3.99 \times 10^{\text{-}13}$ C/m. Upon further sliding by 0.56 {\AA}, the ferroelectric polarization increases to $6.98 \times 10^{\text{-}13}$ C/m. We label this configuration as state I or AB-1 stacking. Conversely, upon a reverse sliding displacement of 1.06 {\AA}, the system transforms into state II or AB-4 stacking. When an additional sliding displacement of 2.31 {\AA} is applied from AB-1, it becomes the AB-2 stacking. Simultaneously, the ferroelectric polarization reduces to $0.59 \times 10^{\text{-}13}$ C/m. More interestingly, the evolution of ferroelectric polarization manifests distinctly in the electronic band structure. As shown in Fig. S5, the band structures near the Fermi level along the $\Gamma$-X path can clearly be observed a coupling from weak to strong, and then decrease. This is completely consistent with the change in the magnitude of the ferroelectric polarization. On the contrary, sliding along the y-axis only modulates the polarization magnitude without changing its directional orientation (see Fig. 2(c)). This further confirms that sliding along the y-axis will not cause ferroelectricity under M$_x$ symmetry.

\begin{figure*}[htb]
\begin{center}
\includegraphics[angle=0,width=1.0\linewidth]{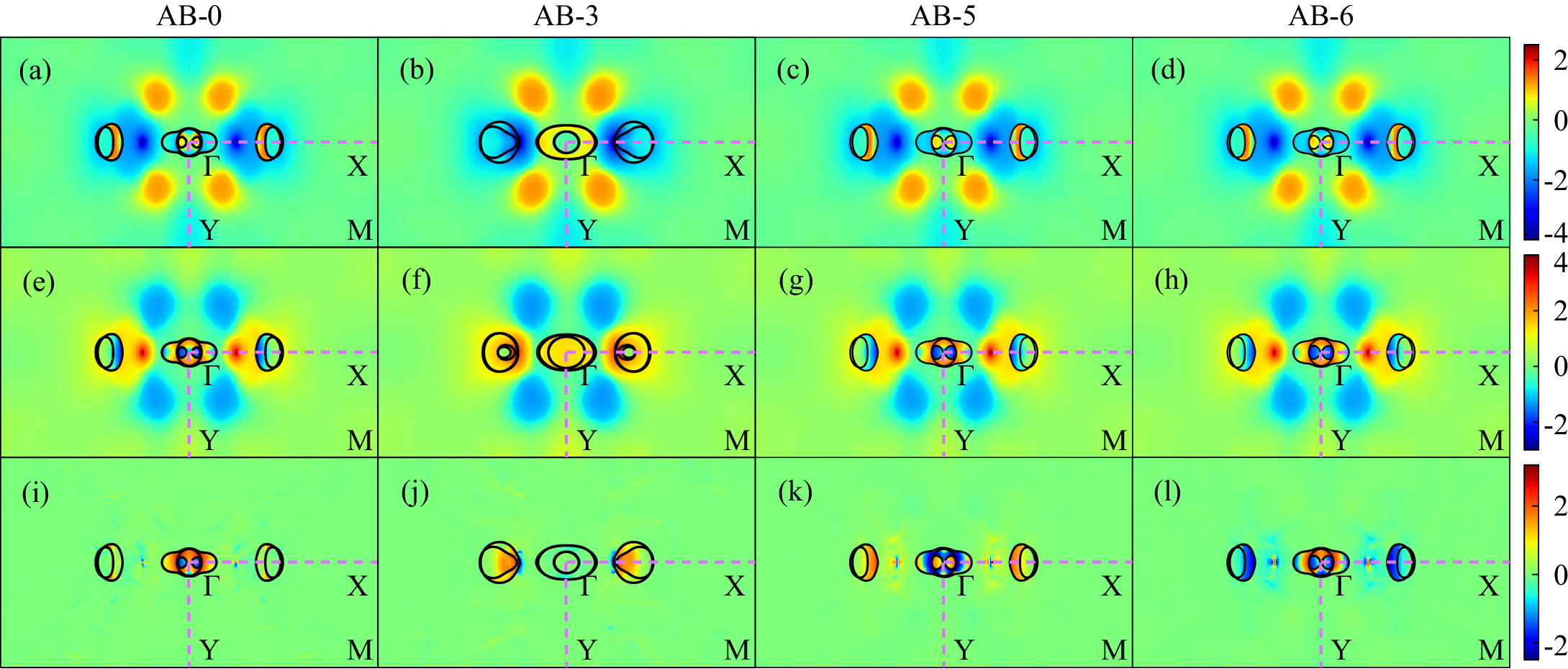}
\caption{The k-resolved SBC on a log scale in 2D BZ slice of k$_z$ = 0 for bilayer 1T'-WTe$_2$. The AB-0, AB-3, AB-5, and AB-6 stacking configurations correspond to (a, e, i), (b, f, j), (c, g, k), and (d, h, l), respectively. (a-d), (e-h), and (i-l) are $\sigma_{xy}^{z}$, $\sigma_{yx}^{z}$ and $\sigma_{yx}^{y}$, respectively.	
}
\end{center}
\end{figure*}

To gain a deeper understanding of the influence of stacking and sliding ferroelectricity on CSHE and ASHE, we systematically calculate the CSHCs and ASHCs of the AA stacking and a series of AB stackings bilayer 1T'-WTe$_2$. These calculation results are summarized in Table I. The CSHCs of bilayer 1T'-WTe$_2$ all increase by nearly two times. Notably, the difference between $\sigma_{xy}^z$ and $\sigma_{yx}^z$ becomes more pronounced and their signs are opposite, further demonstrating the lattice-induced anisotropy. For the ASHCs, the AA stacking and AB stackings show significant differences. The $\sigma_{xy}^y$ and $\sigma_{yx}^y$ in AA stacking exhibit an additive effect of the monolayers, while AB stackings demonstrate a destructive interference effect. When sliding from the paraelectric AB-0 stacking to AB-5 stacking then to the AB-1 stacking along the x-axis, the ASHCs and CSHCs both increase (see Table I). When further sliding to AB-2 stacking, the ASHCs and CSHCs decrease. This indicates that the magnitudes of ASHCs and CSHCs are proportional to the ferroelectric polarization magnitude in the AB stackings. The ferroelectric polarization of the AB-0, AB-1 (AB-4), AB-2, AB-5 (AB-6) stackings are 0.00 C/m, $6.98 \times 10^{\text{-}13}$ C/m ($-6.98 \times 10^{\text{-}13}$ C/m), $0.59 \times 10^{\text{-}13}$ C/m, and $3.99 \times 10^{\text{-}13}$ C/m ($-3.99 \times 10^{\text{-}13}$ C/m), respectively. The CSHCs and ASHCs rule is AB-0 $<$ AB-2 $<$ AB-5 (AB-6) $<$ AB-1 (AB-4), which indicates that the ferroelectric polarization is directly proportional to CSHCs and ASHCs. It means that sliding ferroelectricity can effectively tune CSHCs and ASHCs. It is well known that Pt is a star material for the spin Hall effect. Since there is no breaking of any order parameter, the ASHCs do not exist. We have already confirmed this in our previous studies \cite{11}. Therefore, as long as ASHCs exist, they can in principle be detected in high-quality samples. Experimentally, symmetry reduction is mainly achieved through preparation \cite{13,16,17,18,19}. In 1T'-WTe$_2$, we reduced the symmetry by means of intrinsic order parameters.

It is well known that the band structure varies significantly with the energy level \cite{54,55,56,57,58}. The CSHCs and ASHCs are highly dependent on the band structure. Therefore, understanding the variation of CSHCs and ASHCs as a function of energy level is crucial for the application of high-performance spintronic devices. According to Eq.(1) and Eq.(2), we calculate the CSHCs and ASHCs value varies with the energy level. As shown in Fig. 3 and Fig. S5, the trends of $\sigma_{xy}^{z}$ and $\sigma_{yx}^{z}$ are basically the same. This indicates a high overall similarity in the band structures across the AB stackings series, as also demonstrated in Fig. S6. As shown in Fig. 3(e-h) and Fig. 5(d-f), the ASHCs exhibits significant differences. When the top layer of bilayer 1T'-WTe$_2$ sliding, the ASHC $\sigma_{xy}^{y}$, $\sigma_{yx}^{y}$, and $\sigma_{yy}^{x}$ for the AB-1 stacking are generally larger than those for the other five stackings. The ASHCs values are highly correlated with the magnitude of the ferroelectric polarization. The larger the ferroelectric polarization value of the system, the larger the ASHCs within the energy range of -1 to 1 eV. The maximum value of these ASHC components coincides with the maximum ferroelectric polarization. The ASHC component $\sigma_{xy}^{y}$ of AB-1 stacking reaches a maximum of 141.08 ($\hbar$/e)S/cm at E = E$_F$ + 0.91 eV. The ASHC $\sigma_{yy}^{x}$ of the AB-0 stacking reaches a value of -257.09 ($\hbar$/e)S/cm at E = E$_F$ - 0.69 eV. This large value demonstrates that ASHCs can be realized by tuning the Fermi level via doping or an external electric field. Here is an interesting phenomenon. As shown in Fig. 3 and Fig. S6, we can clearly observe that upon switching the ferroelectric polarization for AB-5 and AB-6 stackings (AB-1 and AB-4 stackings), both the magnitude and direction of the CSHCs remain unchanged. The effect of the change in ferroelectric polarization is reflected in the ASHCs. Although the magnitude of the ASHCs also remains unchanged, their sign is switched. This originates from the fact that the spin polarization direction of spin current of the CSHCs is out-of-plane, whereas that of the ASHCs is in-plane. This further demonstrates that sliding ferroelectricity can effectively modulate the CSHCs and ASHCs. Moreover, Jin $\emph{et al.}$ found that varying the stacking pattern can manipulate the CSHC from positive to negative in bilayer PtTe$_2$ \cite{59}. This change originates from the formation of van der Waals or covalentlike quasibonding between Te atoms in the upper and lower layers. As listed in Table SI, the magnitudes of CSHCs and ASHCs are not strongly correlated with interlayer interactions in 1T'-WTe$_2$. Bilayer PtTe$_2$ and 1T'-WTe$_2$ exhibit entirely distinct mechanisms for manipulating CSHCs.

In addition, we investigate the AA stacking. As shown in Fig. S7, when the SOC is switched off, the enhancement of the coupling effect is mainly manifested near the Fermi level of the $\Gamma$-X path. This is consistent with the fact that the CSHCs and ASHCs of the AA stacking are twice as large as those of a monolayer. Besides, we calculate the $\sigma_{xy}^{z}$ as a function of energy. As shown in Fig. S8, the variation trends of the monolayer and AA stacking bilayer 1T'-WTe$_2$ are significantly different within the energy range of -1 eV to 1 eV. This indicates that the coupling of AA stacking band structure is not simply a simple superposition. Fig. S7(c, d) clearly show an increase in SBC around the $\Gamma$-X path and Y point, which is the origin of the twofold increase in $\sigma_{xy}^{z}$.

To a deeper elucidating the mechanism of sliding ferroelectrically tuning the CSHCs and ASHCs of bilayer 1T'-WTe$_2$, we calculated the band structure and logarithmic-scale k-resolved SBC of $\sigma_{xy}^{z}$, $\sigma_{yx}^{z}$, and $\sigma_{yx}^{y}$. As shown in Fig. 4 and Fig. S9, the contributions of SBC mainly focused on the $\Gamma$-X path and the area around the $\Gamma$ point. We can clearly observe that the SBC signs of CSHC for $\sigma_{xy}^{z}$ and $\sigma_{yx}^{z}$ are opposite. This is consistent with the opposite signs of their CSHC values. For the AB-0, AB-1 (AB-4), AB-2, and AB-5 (AB-6) stackings, the SBC shows little change during the sliding process (see Fig. 4(a-h) and Fig. S9(a-f)). It can also be observed that the SBC of AB-1 (AB-4) stacking is slightly larger than that of AB-5 (AB-6) stacking, and significantly larger than that of AB-2 stacking. However, as shown in Fig. 4(i-l) and Fig. S9(g-i), the interlayer sliding has a very significant impact on the SBC of $\sigma_{yx}^{y}$, which is primarily manifested in the changes around the $\Gamma$-X path and the $\Gamma$ point. It can be clearly observed that AB-1 (AB-4) stacking has the largest SBC, while AB-2 stacking has the smallest. The SBCs of the AB-0 and AB-5 stackings are at intermediate values. This is completely consistent with our calculated results for $\sigma_{yx}^{y}$ rule. Most importantly, we find that the SBC variations of CSHCs for AB-1 and AB-4 (AB-5 and AB-6) stackings are completely identical, while those of $\sigma_{yx}^{y}$ are exactly opposite.

When bilayer 1T'-WTe$_2$ slides from the AB-5 stacking to the AB-6 stacking, the magnitude and sign of the CSHCs remain unchanged, while the magnitude of the ASHCs remains unchanged but their sign is reversed. This originates from the parity of the spin current direction, the charge current direction, and the spin polarization direction in tensor $\sigma^{\gamma}_{\alpha \beta}$. Only when the direction of spin current, charge current, and spin polarization are along the y-axis, it remains unchanged. While along the x-axis and z-axis, it changes sign. Therefore, the CSHCs have even parity, so both their magnitude and sign remain unchanged. On the contrary, the ASHCs have odd parity. Their magnitude remains unchanged while their sign is reversed. Our calculated spin Berry curvature also demonstrates this, as shown in Fig. 5 and Fig. S10.

To understand the variation of the $\sigma_{xy}^{z}$, $\sigma_{yx}^{z}$, and $\sigma_{yx}^{y}$ more comprehensively, we further calculate the k-resolved SBC in the 2D BZ. As shown in Fig. 5(a-h) and Fig. S10(a-f), the difference in SBC between $\sigma_{xy}^{z}$ and $\sigma_{yx}^{z}$ is mainly reflected along the $\Gamma$-X path. It can be observed that the SBC of AB-1 and AB-4 stackings are the most significant. However, the variation in Fig. 5(i-l) is particularly notable. It is mainly manifested as changes near the $\Gamma$ point and along the $\Gamma$-X path. It is straightforward to observe that the magnitude of SBC follows the order AB-1 (AB-4) $>$ AB-5 (AB-6) $>$ AB-0 $>$ AB-2. This is consistent with the magnitude of the polarization (see Fig. 2(b, c)). It implies that sliding ferroelectricity can effectively tune ASHCs.

In summary, we present a strategy to tune the signs and magnitudes of CSHE and ASHE via sliding ferroelectricity. The CSHCs ($\sigma_{xy}^{z}$, $\sigma_{yx}^{z}$) and ASHCs ($\sigma_{xy}^{y}$, $\sigma_{yx}^{y}$) in monolayer 1T'-WTe$_2$ are comparable, both on the order of $\sim$ 50 ($\hbar$/e)S/cm. It is noteworthy that the CSHCs and ASHCs increased by a factor of about two, when a AA bilayer stacking is formed. Moreover, we find that sliding ferroelectricity can effectively tune the CSHCs and ASHCs. The CSHCs and ASHCs are proportional to the magnitude of the ferroelectric polarization. More interestingly, in the ferroelectric switching in AB-1 and AB-4 stackings (AB-5 and AB-6 stackings), the magnitude and sign of the CSHCs remain unchanged, while the magnitude of the ASHCs stays the same and its sign is reversed. We reveal its microscopic mechanism, which the sliding ferroelectricity change the SBCs of the valence and conduction bands around the $\Gamma$-X path. Our work not only provides a way to tune CSHCs and ASHCs, but also an efficient route to realize the field-free switching high-density SOT devices.

\section*{SUPPLEMENTARY MATERIAL}
See the supplementary material for the additional results, including the K-mesh test of 1T'-WTe$_2$ CSHC, band structure of monolayer 1T'-WTe$_2$ without SOC, sliding energy barrier, CSHC and ASHC tensor components as a function of energy, band structure of different AB stackings and AA stacking without SOC, CSHC of AA stacking, SBCs of different AB stackings, and interlayer distance of bilayer 1T'-WTe$_2$.

~\\
\indent This work is supported by the National Natural Science Foundation of China (Grants No. 12474238, and No. 12004295). P. Li also acknowledge supports from the Shaanxi Youth Science and Technology New Star Project (Grant No. 2025ZC-KJXX-71), the China's Postdoctoral Science Foundation funded project (Grant No. 2022M722547), the Fundamental Research Funds for the Central Universities (xxj03202205), and the Open Project of National Laboratory of Solid State Microstructures (M38035), and the Hefei National Research Center for Physical Sciences at the Microscale (KF2025102).

\section*{AUTHOR DECLARATIONS}
$\textbf{Conflict of Interest}$
The authors have no conflicts to disclose.

$\textbf{Author Contributions}$
$\textbf{Chao Wu:}$ Data curation (equal); Investigation (equal); Software (lead). $\textbf{Pengqiang Dong:}$ Investigation (supporting). $\textbf{Kai Wei:}$ Investigation (supporting). $\textbf{Hanbo Sun:}$ Investigation (supporting). $\textbf{Ping Li:}$ Conceptualization (lead); Data curation (equal); Funding acquisition (lead); Investigation (equal); Methodology (lead); Resources (lead); Supervision (lead); Writing-original draft (lead); Writing-review $\&$ editing (lead).

\section*{DATA AVAILABILITY}
The data that support the findings of this study are available from the corresponding authors upon reasonable request.


\bibliography{references}

\end{document}